\documentclass{kapproc} 
\usepackage{psfig}

\normallatexbib

\begin{document}





\articletitle{Spatially resolved spectroscopy of emission-line gas in QSO Host galaxies}

\author{Andrew I. Sheinis}
\affil{UCO/Lick Observatory\\
University of California at Santa Cruz 95064}
\email{sheinis@ucolick.org}

\begin{abstract} We present off-nuclear spectra of 3 radio loud QSO's,
3C249.1, 3C273 and 3C323.1., taken with the echellette spectrograph
and imager (ESI) at Keck observatory. From these spectra we have
extracted the spatial profile along the slit of the [OIII], $\lambda=
\textsf{5007}$ line. Fitted Gaussian distributions to each of these
profiles show emission-line gas out to several tens of kiloparsecs from
the galaxy nucleus. Most observations show several gas components at
distinct velocities and velocity dispersions, much of which is above
the escape velocity for any resonable mass galaxy. In addition, we
show slitless spectroscopy images for one other object, 3C48. From the slitless spactroscopy images we can extract 2-dimensional spatial as well as velocity information on the emission line gas.
\end{abstract}

\section{Spatially resolved Off-Nuclear spectroscopy}

We have taken off-nuclear spectra of 3 radio loud QSO's, 3C249.1,
3C273 and 3C323.1 (Boroson and Oke 1984), using the Echellette
spectrograph and imager (ESI) (Sheinis et. al 2000) at Keck
observatory.  Each object was observed at several differant position
angles and offsets from the nucleus. From these spectra we have
extracted the spatial profile along the slit of the [OIII], $\lambda=
\textsf{5007}$ line.

Each image is the median of four 15 minute exposures. The data were
processed using the Information Data Language (IDL). They were first
rectified to remove the instrument distortion, bias subtracted then a
two dimensional sky model was subtracted. After this extraction we fit a
linear combination of two Gaussian distributions to each of these
profiles, using (IDL).

Figure 1  shows the two-dimensional extraction image for each object. For 3C249.1, three slit position are shown, 3 seconds east, 3.5 seconds north and 3 seconds west are shown. One position is shown for each of the remaining objects, 3C273 and 3C323.1. Those positions are 4 seconds east and 3 seconds east. The lower panels of figure 1 show the results of the Gaussian fit for each object, namely the mean, dispersion and magnitude of the Gaussian fit for each gas component.

The plots show bright extended emission at tens of kiloparsecs from
the galaxy nucleus. This emission is observed to have peak velocity
$\textsf{FWHM} =\textsf{2.35 x }\sigma$ of 540, 940 and 535 km/sec for
the three slit positions of 3C249.1, 500 km/sec for 3C273 and 470
km/sec for 3C323.1. These velocities are most likely well beyond
escape velocity for these galaxies.



 \hspace{\fill}
  \noindent\begin{minipage}{5.0in}
  \vspace{0.2in}
   \psfig{file=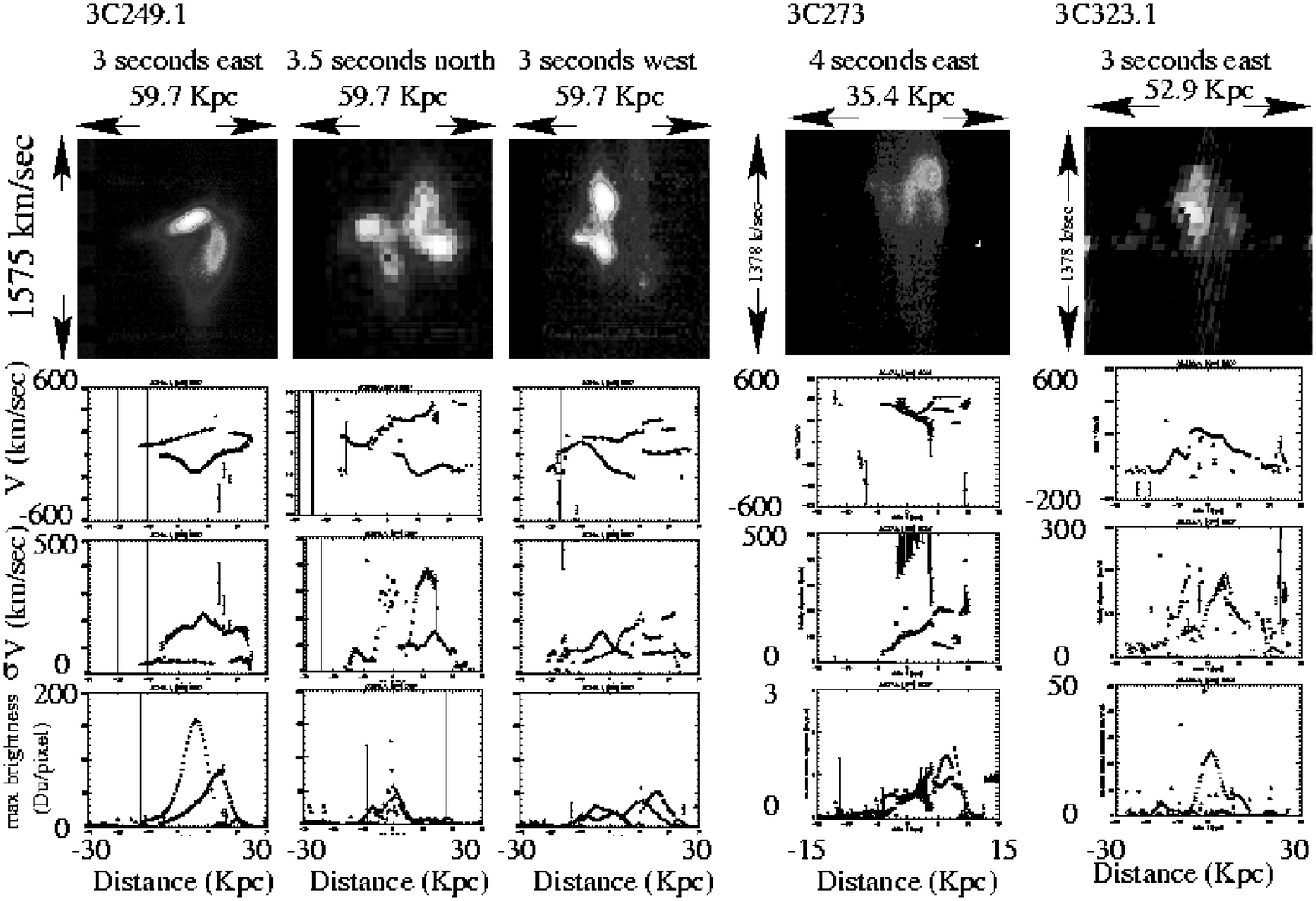,width=5.25in,angle=0.0}
  \refstepcounter{figure}
  \centerline{\textbf{Figure \thefigure.}3C249.1, 3C273 and 3C323.1  seen in [OIII], spatially resolved.}
 \label{3c323.1}
 \end{minipage}

\vspace{0.2in}

\section{Slitless Spectroscopy}

We have obtained slitless spectroscopy images of several objects. The
four panels of figure 2 show ten minute exposures of one well-studied
object, 3C48 (Canalizo 2000) taken at four differant position angles,
through a six arcsecond wide slit. They have been processed
identically to the above images. Additionally, a modelled QSO
continuum was carefully subtracted to reveal details within a half
arcsecond of the nucleus.

In order to deconvolve the velocity and spatial information in the y
axis of each image, we have produced pairs of images that have the
dispersion axis rotated relative each other, by 180 degrees. In
objects (like 3C48) that contain well defined clumps of gas moving at
a definite velocity, position and velocity can be inferred directly by
comparing the two images. In objects with a more complex spatial and
velocity structure this method becomes less effective.



 \hspace{\fill}
  \noindent\begin{minipage}{5.0in}
  \vspace{0.2in}
   \psfig{file=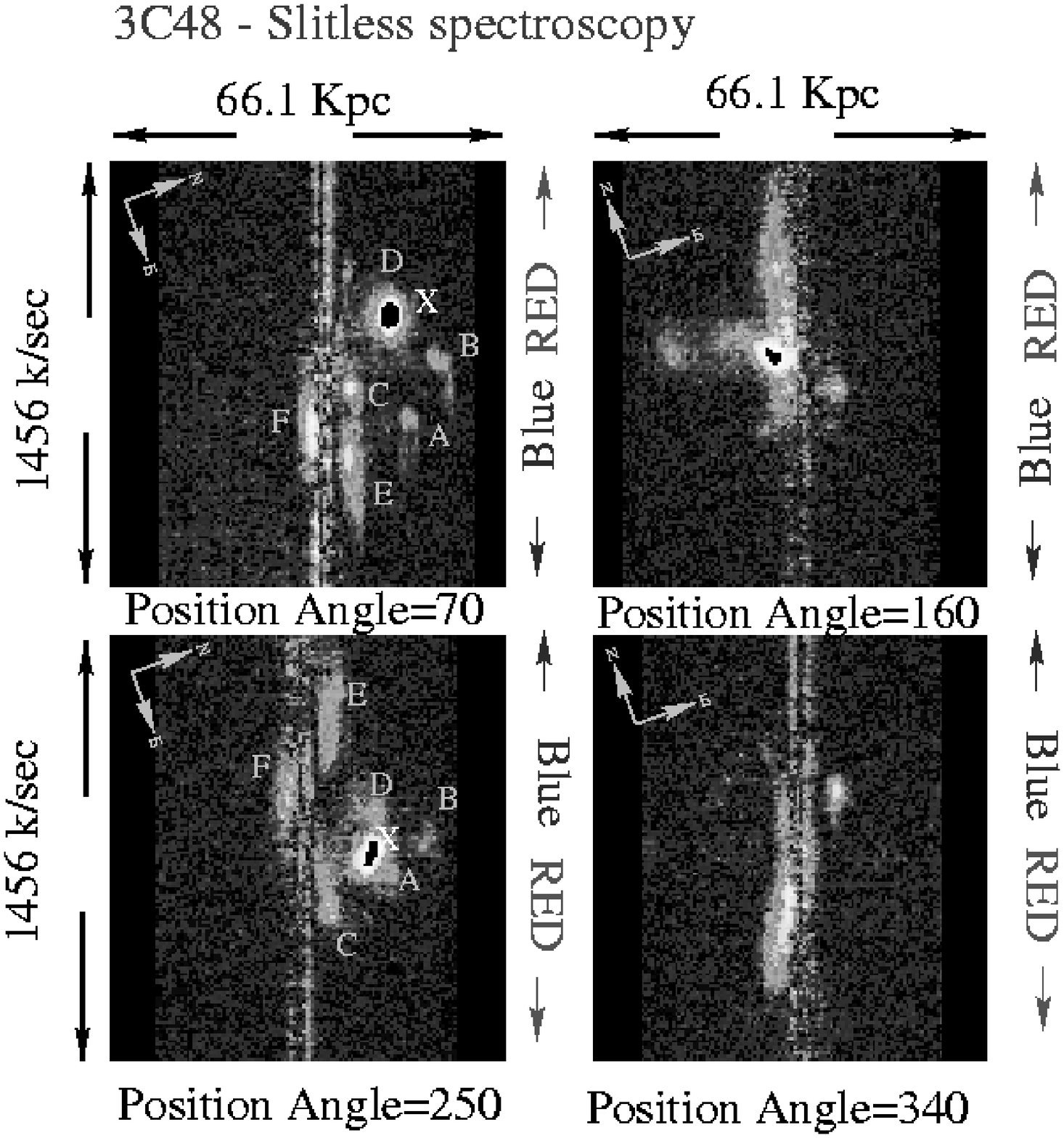,width=3.25in,angle=0.0}
  \refstepcounter{figure}
  \centerline{\textbf{Figure \thefigure.}Slitless spectroscopy of 3C48,  seen in [OIII]. }
 \label{3C48}
 \end{minipage}
  \vspace{0.2in}

Figure 2 shows the emission-line gas of 3C48 as viewed through a 6
arcsecond wide by 20 arcsecond long slit. The two images show two
differant position angles separated by 180 degrees. In first pair of
images of 3C48 (upper and lower left) we see one bright gas knot (X)
redshifted by $\approx$ 200 km/sec relative to four smaller knots
(A-D) that appear to be at similar velocites. All 5 knots are located
to the north of the nucleus. Closer to the nucleus we see one higher
dispersion ($\approx$ 600 km/sec), higher velocity knot ($\approx$
500km/sec) (E) to the north and another (F) to the south. A similar
evaluation can be done to the rightmost pair of images.

This method has shown promise to produce spatial and velocity
information in emission-line gas.



\end{document}